\documentclass[12pt,letterpaper]{article}
\usepackage{setspace}
\usepackage[margin=1in]{geometry}
\usepackage[colorlinks,citecolor=blue]{hyperref}
\usepackage{mathrsfs}
\usepackage{bm, physics}
\usepackage{color}
\usepackage{caption}
\usepackage{natbib}

\usepackage{graphicx}
\usepackage{algorithm}
\renewcommand{\hat}{\widehat}
\renewcommand{\bar}{\overline}

\usepackage{algpseudocode}

\def\m{\mathcal}

\begin{document}

\title{Bayesian Group Learning for Shot Selection of
Professional Basketball Players
}

\author{Guanyu Hu~~~~Hou-Cheng Yang~~~~Yishu Xue}
\maketitle

\abstract{
In this paper, we develop a group learning approach to
analyze the underlying heterogeneity structure of shot selection among
professional basketball players in the NBA. We propose a mixture of finite
mixtures (MFM) model to capture the heterogeneity of shot selection among
different players based on Log Gaussian Cox process (LGCP). Our proposed method
can simultaneously estimate the number of groups and group configurations. An
efficient Markov Chain Monte Carlo (MCMC) algorithm is developed for our
proposed model. Simulation studies have been conducted to demonstrate its
performance. Ultimately, our proposed learning approach is further
illustrated in analyzing shot charts of selected players in the NBA's 2017--2018
regular season.

\noindent
Keywords: 
Basketball Shot Charts; Heterogeneity Pursuit; Log Gaussian Cox Process;
Mixture of Finite Mixtures; Nonparameteric Bayesian}

\section{Introduction}
In basketball data analytics, one primary problem of research interest is to
study how players choose the locations to make shots. Shot charts, which are
graphical representations of players' shot location selections, provide
important summary of information for basketball coaches as well as teams' data
analysts, as no good defense strategies can be made without understanding the
shot selection habits of players in the rival teams. Shot selection data have
been discussed from different statistical perspectives. \cite{reich2006spatial}
developed a spatially varying coefficients model for shot-chart data, where the
court is divided into small regions, and the probability of making a shot in
these zones are modeled using the multinomial logit approach. Recognizing the
random nature of shot location selection, \cite{miller2014factorized} analyzed
the underlying spatial structure among professional basketball players based on
spatial point processes. \cite{franks2015characterizing} combined spatial and
spatio-temporal processes, matrix factorization techniques and hierarchical
regression models for characterizing the spatial structure of locations for shot
attempts. In spatial point processes, locations for points are assumed random,
and are regarded as realizations of a process governed by an underlying
intensity. Spatial point processes are well discussed in many statistical
literatures, such as the Poisson process \citep{geyer1998likelihood}, the Gibbs
process \citep{goulard1996parameter}, and the Log Gaussian Cox process
\citep[LGCP;][]{moller1998log}. In addition, they have been applied to different
areas, such as ecological studies
\citep{thurman2015regularized,jiao2020heterogeneity}, environmental sciences
\citep{veen2006assessing,hu2019new}, and sports analytics
\citep{miller2014factorized,jiao2019bayesian}. Most existing literatures
concentrate on parametric \citep{guan2008consistent} or nonparametric
\citep{guan2008consistent,geng2019bayesian} estimation of the underlying
intensities for spatial point process and analysis of second-order properties
\citep{diggle2007second}. There are very limited literatures discussing the
grouping pattern of multiple point processes. Knowing the group information of
different point processes will lead to discovery of the underlying heterogeneity
structure of different players.

\cite{jiao2019bayesian} proposed a joint model approach for basketball shot
chart data. After model parameter estimates are obtained for different players,
they are grouped via \emph{ad hoc} clustering approaches, such as hierarchical
clustering. \citet{chen2019case} developed a group linked Cox process model for
analyzing point of interest (POI) data in Beijing. To determine the number of
groups, starting from the most complicated model where each observation has its
own group, a loss function is used in a series of hierarchical merging steps to
combine the groups. In both methods, the inherent uncertainty in estimation for
the number of groups is ignored. In contrast, Bayesian models such as the
Dirichlet process \citep[DP;][]{ferguson1973bayesian} offer a natural solution
that simultaneously estimates the number of groups and group configurations.
However, \citet{miller2013simple} shows that Dirichlet process mixture model
(DPMM) tends to create tiny extraneous groups. In other words, DPMM does not
produce consistent estimator of the number of groups. In this paper, we employ
the mixture of finite mixture \citep[MFM;][]{miller2018mixture} approach for
learning the group structure of multiple spatial point processes, which, on the
contrary, provides consistent estimation for group numbers.

The contribution of this paper is two-fold. First, we propose a Bayesian group
learning method to simultaneously estimate the number of groups and the group
configurations. In particular, we use an LGCP to model the spatial pattern of
the shot attempts. Based on similarity matrices of fitted intensity among
different players, a MFM model is incorporated for group learning. Moreover, the
MFM model has a P\'{o}lya urn scheme similar to the Chinese restaurant process,
which is exploited to develop an efficient Markov chain Monte Carlo MCMC
algorithm without reversible jump or even allocation samplers. Compared with
existing approaches \citep[e.g.,][]{jiao2019bayesian}, our proposed method does
not require prior information for the number of groups, and grouping is
incorporated into the structure of the model and performed directly based on
shot selection intensity instead of via \emph{ad hoc} analysis of regression
coefficients. In addition, our proposed Bayesian approach reveals interesting
shooting patterns of professional basketball players, and the summaries better
characterize player types beyond the traditional position categorization.

The rest of the paper is organized as follows. In Section~\ref{sec:data}, the
shot chart data of different players from the 2017--2018 NBA regular season is
introduced. In Section~\ref{sec:method}, we discuss the LGCP and develop the
Bayesian group learning method based on MFM. Details of the Bayesian inference
are presented in Section~\ref{sec:bayesian_inference}, including the MCMC
algorithm and post MCMC inference methods. Simulation studies are conducted in
Section~\ref{sec:simu}. Applications of the proposed methods to NBA players data
are reported in Section~\ref{sec:real_data}. Section~\ref{sec:discussion}
concludes the paper with a discussion.

\section{Motivating Data}\label{sec:data}

Our data consists of both made and missed field goal attempt locations from the
offensive half court of games in the 2017--2018 National Basketball Association
(NBA) regular season. The data is available
at~\url{http://nbasavant.com/index.php}. We focus on players that have made more
than 400 field goal attempts (FTA). Also, players who just started their careers
in the 2017--2018 season, such as Lonzo Ball and Jayson Tatum, are not
considered. A total of 191 plays who meet the two criteria above are included in
our analysis.

\begin{figure}[tbp]
	\centering
	\includegraphics[width=\textwidth]{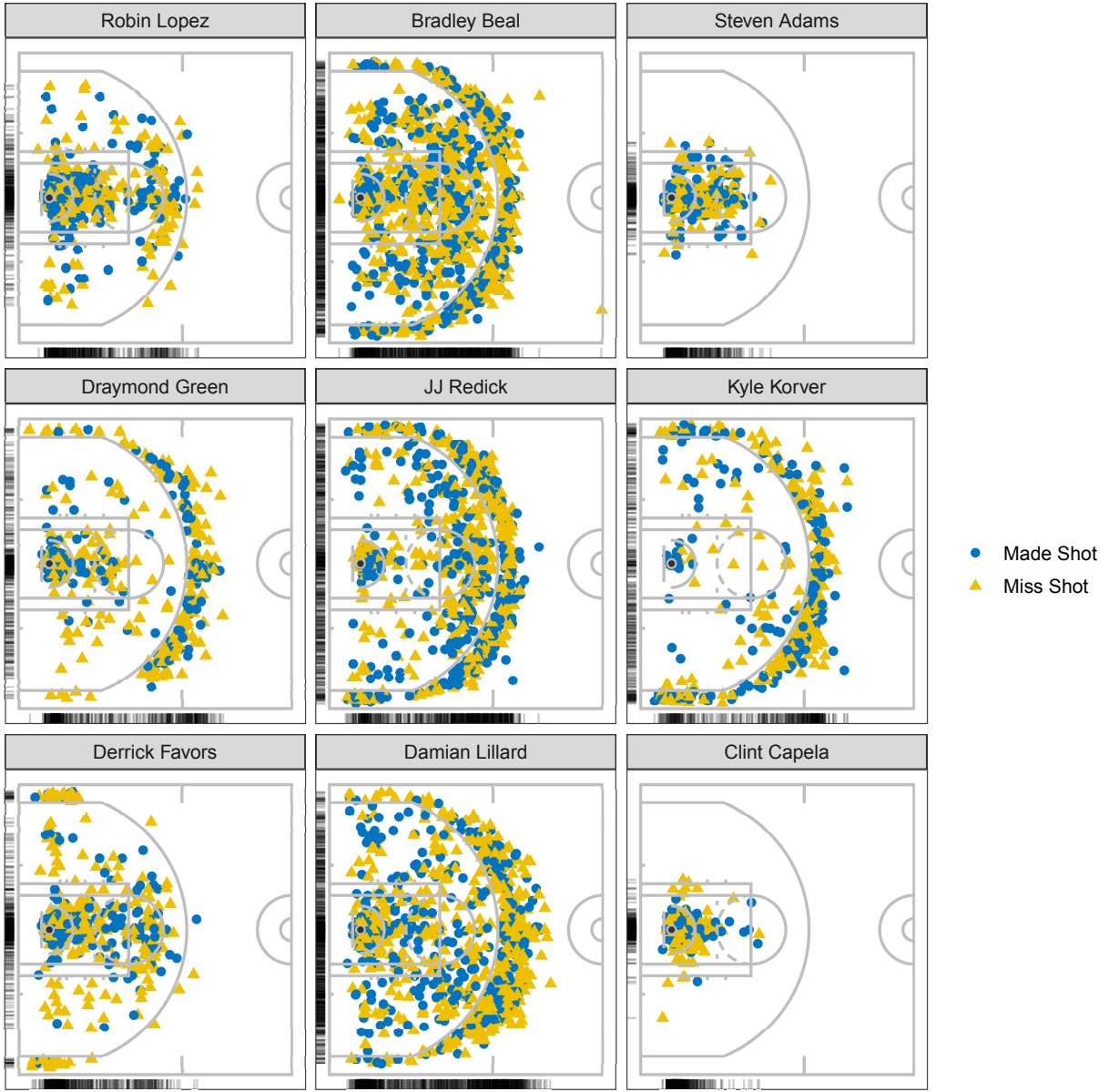}
	\caption{Shot charts for selected NBA players.}
	\label{fig:shotcharts9}
\end{figure}

We model a player's shooting location choices as a spatial point pattern on the
offensive half court, a~47 ft by~50 ft rectangle, which is the standard size for
NBA. We assume the spatial domain $D\in[0,47]\times[0,50]$. Indexing the players
with $i\in\{1,\ldots, 191\}$, the locations of shots, both made and missed, for
player~$i$ are denoted as $X_i = \{x_{i,1},\ldots, x_{i,T_i}\}$, $\forall
x_{i,T_i}\in D$, where~$T_i$ is the total number of attempts made by player~$i$
on the offensive half court. We select nine players and visualize their shot
charts in Figure~\ref{fig:shotcharts9}. It can be seen that Clint Capela makes
more shot attempts in the painted area, as most of his field goals are slam
dunks. JJ Redick, however, prefers to shoot outside the painted area. Our goal
is to find groups of similar shooting location habits among the NBA basketball
players.

\section{Method}\label{sec:method}

\subsection{Log-Gaussian Cox Process}

The shot locations can be denoted as $y = (\bm{s}_1,\ldots, \bm{s}_\ell)$, with
$(\bm{s}_1,\ldots, \bm{s}_\ell)$ being the locations of points that are observed
within a bounded region~$\mathcal{B}\subseteq \mathcal{R}^2$. Such spatial point
pattern can be regarded as a realization from a spatial point process $Y$.
Spatial point pattern data are modeled by spatial point processes
\citep{diggle1976statistical} characterized by a quantity called intensity.
Within a region~$\mathcal{B}$, the intensity on any location~$\bm{s} \in
\mathcal{B}$ can be represented as~$\lambda(\bm{s})$, which is defined as:
\begin{equation*}
\lambda(\bm{s}) = \lim_{|\dd  \bm{s}\rightarrow
    0|}\left(\frac{\textrm{E}[N(\dd \bm{s})]}{|\dd  \bm{s}|} \right),
\end{equation*}
where~$\dd \bm{s}$ is an infinitesimal region around~$\bm{s}$, $|\dd \bm{s}|$
represents its area, and $N(\dd \bm{s})$ shows the number of events that
happened over $\dd  \bm{s}$. For an area~$A\subseteq \mathcal{B}$, we denote by
$N_{\bm{Y}}(A)=\sum_{i=1}^\ell 1(s_i\in A)$ the counting process associated with
the spatial point process $\bm{Y}$, which counts the number of points in a
realization of $\bm{Y}$ that fall within an area $A\subseteq\mathcal{B}$. The
Poisson distribution has been conventionally used for modeling count data, and
correspondingly, the spatial Poisson point process is a popular tool for
modeling spatial point pattern data. For a Poisson process $Y$
over~$\mathcal{B}$, with its intensity function denoted as $\lambda(\bm{s})$,
the counting process $N_Y(A)$ satisfies:
\begin{equation*}
N_Y(A) \sim \mbox{Poisson}(\lambda(A)),\mbox{ where }\lambda(A) = \int_A 
\lambda(\bm{s})\dd \bm{s}.
\end{equation*}
For the Poisson process, it is easy to obtain
$E(N_{\bm{Y}}(A))=\text{Var}(N_{\bm{Y}}(A))=\lambda(A)$. When
$\lambda(\bm{s})=\lambda$, we have constant intensity over the space
$\mathcal{B}$ and in this special case, $\bm{Y}$ reduces to a homogeneous
Poisson process (HPP). For a more general case, $\lambda(\bm{s})$ can be
spatially varying, which leads to a nonhomogeneous Poisson process (NHPP). For
the NHPP, the log-likelihood on $\mathcal{B}$ for the observed dataset $\bm{y}$
is given by
\begin{align}
	\ell=\sum_{i=1}^k \text{log}\lambda(s_i)-\int_{\mathcal{B}} \lambda(s) \dd s,
	\label{pploglike}
\end{align}
where $\lambda(s_i)$ is the intensity function for location $s_i$.
We signify that a set of points $\bm{y}=(s_1,s_2,...,s_\ell)$ follows a Poisson
process as
\begin{equation}
	\bm{y}\sim \mathcal{PP}(\lambda(\cdot)).
	\label{eq:spp}
\end{equation}
A log-Gaussian Cox process (LGCP) is a doubly-stochastic Poisson process with a
spatially varying intensity function modeled as an exponentiated Gaussian
process, i.e., Gaussian random field \citep[GRF;][]{rasmussen2003gaussian},
which is a spatially continuous random process in which random variables at any
location in a space are normally distributed, and are correlated with random
variables at other locations according to a continuous correlation process. The
LGCP can be written hierarchically as
\begin{equation}
\label{eq:h_lgcp}
	\begin{split}
	\bm{y} &\sim \mathcal{PP}(\lambda(\cdot)),\\
	\lambda(\cdot) &=\exp(Z(\cdot)),\\
	Z(\cdot)& \sim \mathcal{GP}(0,k(\cdot,\cdot)),	
	\end{split}
\end{equation} 
where $k(\cdot,\cdot)$ is the covariance function of the Gaussian process,
$Z(\cdot)$. For estimation, the GRF is approximated by the solution to a
stochastic partial differential equation \citep[SPDE; see,][ for a
review]{lindgren2011explicit}, as SPDEs provide an efficient way of
approximating the GRF in continuous space \citep{simpson2016going}. Under a
purely Bayesian paradigm, model-based Markov chain Monte Carlo (MCMC) can be
time-consuming for LGCP. Therefore, we compute the LGCP using the integrated
nested Laplace approximation \citep[INLA;][]{rue2009approximate}, which is an
alternative to MCMC for fitting latent Gaussian models that provides a fast and
accurate way to fit a potential model, and facilitates computationally efficient
inference on point processes. For more details about INLA, we refer the reader to
the R-INLA project website at~\url{http://www.r-inla.org}. Details about
computation is directed to Section~\ref{sec:bayesian_inference}.

With estimated intensity surfaces for $n$ different players, denoted as
$\hat{\lambda}^{(1)}(\cdot),\ldots, \hat{\lambda}^{(n)}(\cdot)$. With our main
goal being to group players who share similar shot location choices over the
court, an appropriate metric is needed to quantify similarities among the
intensities. Let the matrix~$\textbf{C}$ be that
$\textbf{C}\equiv(\hat{\lambda}^{(1)}, \hat{\lambda}^{(2)}, \ldots,
\hat{\lambda}^{(n)})$, and denote $\textbf{C}^{(i)}=\lambda^{(i)}$. Then,
following the approach in \cite{cervone2016multiresolution}, we compute the
players' similarity matrix $\textbf{H}$ as:
\begin{equation}
\text{H}_{i,j}= \exp\left\{-\left\Vert
\frac{\textbf{C}^{(i)}}{\sum_{}^{}\textbf{C}^{(i)}}
- \frac{\textbf{C}^{(j)}}{\sum_{}^{}\textbf{C}^{(j)}}\right\Vert\right\},
\label{eq:k_estimation}
\end{equation}
where $i,j\in \{1,\ldots, n\}$ and $\left\Vert\cdot\right\Vert$ is $L_2$ norm.
It can be seen that $\textbf{H}$ is symmetric, and $\textbf{H}\in
\mathcal{R}^{n\times n}$.

\subsection{Group Learning via Point Process Intensity}
With the similarity matrix $\textbf{H}$ obtained, we employ nonparametric
Bayesian methods to detect grouped patterns in the intensities. Our initial step
is to transform the similarity matrix $\textbf{H}$ so that each entry $H_{i,j}$
is within the range of a Gaussian distribution. Denote the fisher transformed
distance matrix \citep{fisher1915frequency} as~$\bm{\mathscr{\bm{S}}}$. Its
$(i,j)$th element is calculated as
\begin{equation}\label{eq:fisherZ}
\mathscr{S}_{ij}  = \frac{1}{2}\log\left(\dfrac{1+\text{H}_{i,j}}
{1 - \text{H}_{i,j}}\right),
\end{equation}
A larger value of $\mathscr{S}_{ij}$ indicate higher similarity of intensities.
We further assume that
\begin{equation}\label{eq:sBm}
\begin{split}
\mathscr{S}_{ij} &\mid \bm{\mu}, \bm{\tau}, k \sim
\mbox{N}(\mu_{ij},\tau_{ij}^{-1}),
\quad \mu_{ij} = U_{z_i z_j}\\
\tau_{ij} &= T_{z_i z_j}, \quad 1 \leq i < j
\leq n,
\end{split} 
\end{equation}
where~$k$ denotes the number of groups, $\mbox{N}()$ denotes the normal
distribution, $z_i\in\{1,\ldots,k\}$ denotes the group membership of player~$i$
for~$i=1,\ldots, 191$. The matrices $\bm{U} = [U_{rs}] \in (-\infty,+\infty)^{k
\times k}$ and $\bm{T} = [T_{rs}] \in (0,+\infty)^{k \times k}$ are both
symmetric, with $U_{rs} = U_{sr}$ indicating the mean closeness between any two
fitted intensity surfaces in groups $r$ and $s$, respectively, and~$T_{rs} =
T_{sr}$ indicating the precision.

Denote by $\m Z_{n, k} = \big\{(z_1, \ldots, z_n) : z_i \in \{1, \ldots, k\}, 1
\le i \le n \big\}$ the set of all possible partitions of $n$ players into $k$
groups. With certain $z \in \m Z_{n, k}$, denote by $\mathscr{S}_{[rs]}$ the
$n_r \times n_s$ sub-matrix of $\mathscr{S}$ consisting of entries
$\mathscr{S}_{ij}$ where $z_i = r$ and $z_j = s$. Under model~\eqref{eq:sBm},
the joint likelihood of $\mathscr{S}$ can be written as
\begin{equation}\label{eq:like}
\begin{split}
P(\bm{\mathscr{S}} \mid \bm{z}, \bm{U}, \bm{T}, k) = \prod _{1 \leq r\leq s
\leq k}
P(\bm{\mathscr{S}}_{[rs]}\mid
\bm{z}, \bm{U},\bm{T}),
\end{split}
\end{equation}
where
\begin{equation*}
\begin{split}
P(\bm{\mathscr{S}}_{[rs]} \mid \bm{z}, \bm{U},\bm{T}) = \prod_{1\leq i < j
\leq n:
z_i = r, z_j = s} \frac{1}{\sqrt{2\pi T_{rs}^{-1}}}
\exp\left\{-\frac{T_{rs}(\mathscr{S}_{ij}-U_{rs})^2}{2}\right\}.
\end{split}
\end{equation*}
Assuming that the number of groups $k$ is given, independent prior distributions
are often assigned to $z$, $U$,  and~$T$. Such specification can be conveniently
incorporated into a finite mixture model. When $k$ is unknown, however, the
Dirichlet process mixture prior models \citep{antoniak1974mixtures} can be
employe as:
\begin{eqnarray}\label{eq:DPMM}
\bm{\mathscr{S}}_i \sim   F(.,\bm{\theta}_i), \quad \bm{\theta}_i \sim  G(.),
\quad G \sim  DP(\alpha G_0),
\end{eqnarray}
with $\bm{\mathscr{S}_i} =
(\mathscr{S}_{i1},\mathscr{S}_{i2},\ldots,\mathscr{S}_{in})$, $\bm{\theta}_i =
(\bm{\theta}_{i1},\bm{\theta}_{i2},\ldots,\bm{\theta}_{in})$ and
$\bm{\theta}_{ij} = (\mu_{ij},\tau_{ij})$. The process $G$ is parameterized by a
base measure $G_0$ and a concentration parameter $\alpha$. With $\bm{\theta}_i$
for $i=1,\ldots, n$ drawn from $G$, a conditional prior distribution for a newly
drawn $\bm{\theta}_{n+1}$ can be obtained via integration
\citep{blackwell1973ferguson}:
\begin{eqnarray}\label{eq:DPMM1}
p(\bm{\theta}_{n+1}\mid \bm{\theta}_1,\ldots,\bm{\theta}_n) =
\dfrac{1}{n+\alpha}\sum_{i=1}^n\delta_{\bm{\theta}_i}(\bm{\theta}_{n+1}) +
\dfrac{\alpha}{n+\alpha}G_0(\bm{\theta}_{n+1}),
\end{eqnarray}
with $\delta_{\bm{\theta}_i}(\bm{\theta}_{j}) = I(\bm{\theta}_j =
\bm{\theta}_i)$ being the point mass at $\bm{\theta}_i$. The model can be
equivalently obtained with the introduction of group membership $z_i$'s and
having $K$, the number of groups, approach infinity \citep{neal2000markov}:
\begin{equation}\label{eq:DPMM2}
\begin{split}
\bm{\mathscr{S}}_i \mid z_i, \bm{\theta}^* & \sim F(\bm{\theta}^*_{z_i}),\\
z_i \mid  \bm{\pi} & \sim \text{Discrete} (\pi_1,\ldots,\pi_K), \\
\bm{\theta}^*_c & \sim G_0,\\
\bm{\pi} & \sim \text{Dirichlet}(\alpha/K,\ldots ,\alpha/K) ,
\end{split}
\end{equation}
where $\bm{\pi} = (\pi_1,\ldots,\pi_K)$. It can be seen that under this
construction, the group-specific distribution $F(\cdot \mid \bm{\theta}_c^*)$
solely depends on the vector of parameters $\bm{\theta}^*_c$.

In construction~\eqref{eq:DPMM2}, the prior distribution of $(z_1,\ldots,z_n)$,
which would allow for automatic inference on the number of groups $k$, can be
obtained by integrating out $\bm{\pi}$, the mixing proportions. This is also
known as the Chinese restaurant process 
\citep
[CRP;][]{aldous1985exchangeability,pitman1995exchangeable,neal2000markov}.
The conditional distribution for $z_i$ is defined through the metaphor of a
Chinese restaurant \citep{blackwell1973ferguson}:
\begin{eqnarray}\label{eq:crp}
P(z_{i} = c \mid z_{1}, \ldots, z_{i-1})  \propto   
\begin{cases}
\vert c\vert  , &  \text{at an existing table labeled}\, c\\
\alpha,  & \text{if} \, $c$\,\text{is a new table}
\end{cases},
\end{eqnarray}
where $\vert c\vert$ denotes the size of group $c$. Despite its ability to
simultaneously estimate the number of groups and group configuration, the CRP
has been shown by \cite{miller2018mixture} to produce redundant tail groups,
causing inconsistency in estimation for the number of groups even with the
sample size going to infinity. \cite{miller2018mixture} also proposed a
modification of the CRP, known as the mixture of finite mixtures (MFM) model, to
mitigate this problem. The MFM model can be formulated as:
\begin{eqnarray}\label{eq:MFM}
k \sim p(\cdot), \quad (\pi_1, \ldots, \pi_k) \mid k \sim \mbox{Dirichlet}
(\gamma,
\ldots, \gamma), \quad z_i \mid k, \bm{\pi} \sim \sum_{h=1}^k  \pi_h
\delta_h,\quad
i=1, \ldots, n, 
\end{eqnarray}
with $p(\cdot)$ being a proper probability mass function on the set of positive
integers, and $\delta_h$ being a point mass at $h$. Define a
coefficient~$V_n(w)$ as
\begin{equation*}
	V_n(w) = \sum_{k=1}^{+\infty}\dfrac{k_{(w)}}{(\gamma k)^{(n)}} p(k),
\end{equation*}
where $w$ denotes the number of ``existing tables'',
$k_{(w)}=k(k-1)\ldots(k-w+1)$, and $(\gamma k)^{(n)} = {\gamma k}(\gamma
k+1)\ldots(\gamma k+n-1)$, $x^{(0)} = 1$, and~$x_{(0)}=1$. Introduction of a
new
table is slowed down by $V_n(w+1) / V_n(w)$, which yields the following
conditional prior of $\bm{\theta}$:
\begin{eqnarray}\label{eq:MFM1}
P(\bm{\theta}_{n+1}\mid \bm{\theta}_1,\ldots,\bm{\theta}_n) \propto
\sum_{i=1}^w(n_i + \gamma)\delta_{\bm{\theta}^*_i} +
\dfrac{V_n(w+1)}{V_n(w)}\gamma G_0(\bm{\theta}_{n+1}),
\end{eqnarray}
with $\bm{\theta}^*_1,\dots,\bm{\theta}^*_w$ being the unique values taken by
$\bm{\theta}_1,\ldots, \bm{\theta}_n$. Conditional distribution for the group
membership can be expressed analogous
to~\eqref{eq:crp}~as:
\begin{eqnarray}\label{eq:mcrp}
P(z_{i} = c \mid z_{1}, \ldots, z_{i-1})  \propto   
\begin{cases}
	\vert c\vert + \gamma  , &  \text{at an existing table labeled }\, c\\
V_n(w+1)/ V_n(w)\gamma,  & \text{if } \, c\,\text{ is a new
table}
\end{cases}.
\end{eqnarray}
Adapting MFM to our model setting for functional grouping, the model and prior
can be expressed hierarchically as: 
\begin{align}\label{eq:Func_MFM}
& k \sim p(\cdot), \text{where $p(\cdot)$ is a p.m.f on \{1,2, \ldots\} }
\nonumber \\
& T_{rs} = T_{sr} \stackrel{\text{ind}} \sim \mbox{Gamma}(\alpha, \beta),  
\quad r, s = 1, \ldots, k,  \nonumber
\\ \nonumber
& U_{rs} = U_{sr} \stackrel{\text{ind}} \sim \mbox{N}(\mu_0, k_0^{-1} 
T_{rs}^{-1}),  \quad r, s = 1, \ldots, k,  \nonumber
\\
& \mbox{pr}(z_i = j \mid \bm{\pi}, k) = \pi_j, \quad j = 1, \ldots, k, \, i =
1,
\ldots, n,  \\ \nonumber
& \bm{\pi} \mid k \sim \mbox{Dirichlet}(\gamma, \ldots, \gamma),\\
& \mathscr{\bm{S}}_{ij} \mid \bm{z}, \bm{U},\bm{T}, k \stackrel{\text{ind}}
\sim
\mbox{N}(\mu_{ij}, 
\tau_{ij}^{-1}), \quad \mu_{ij} = U_{z_i z_j},~~ \tau_{ij} = T_{z_i z_j},~~ 1
\leq i < j \leq n.  \nonumber
\end{align}
We assume $p(\cdot)$ is a $\mbox{Poisson}(1)$ distribution truncated to be
positive through the rest of the paper, which has been proved by
\cite{miller2018mixture,geng2019probabilistic} to guarantee consistency for the
mixing distribution and the number of groups.  We refer to the hierarchical
model above as MFM-PPGrouping. 

\section{Bayesian Inference}\label{sec:bayesian_inference}

\begin{figure}[tbp]
	\centering
	\includegraphics[width=0.6\textwidth]{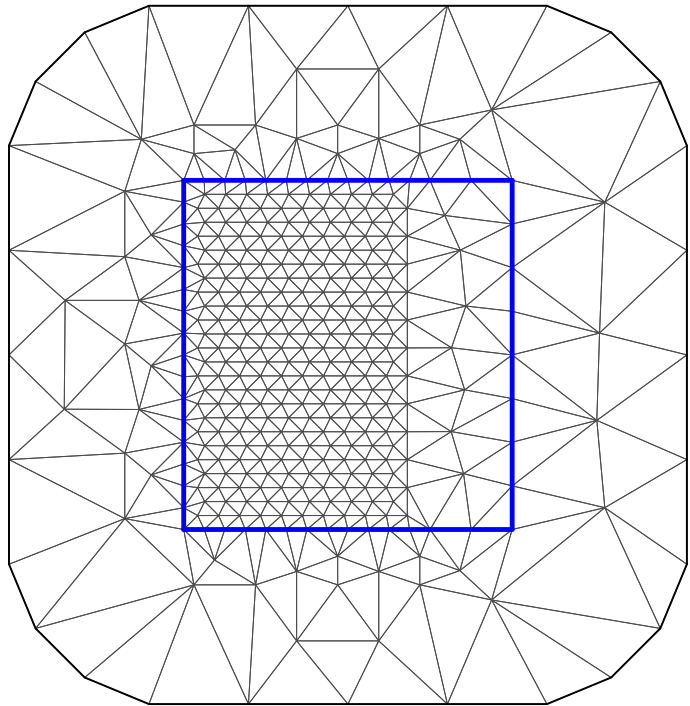}
\caption{Triangulation for the shot data locations over which the ``tent"
functions are constructed (black line), and the observation locations are inside
the blue rectangle.}
	\label{mesh}
\end{figure}

In this section, we discuss the implementation of INLA estimation for LGCP,
collapsed sampler algorithm for the proposed MFM-PPGrouping approach, and
posterior inference on MCMC samples. INLA tries to partition a region to
disjoint triangles (i.e., triangulation), and uses this mesh of discrete
sampling locations to estimate a continuous surface in space via interpolation.
A set of piecewise linear basis functions, which are typically ``tent'', or
finite element functions, are defined over a triangulation of the domain of
interest. The mesh is composed of two regions: the interior mesh, which is where
the actions happen; and the exterior mesh, which is designed to alleviate the
boundary effects. It is formed by partitioning the region into triangles. The
more triangles we have, the more precise is our approximation, at the cost of
extended computational time. And the desired mesh would have small triangles
where the shot data is dense, larger where the shot data is more sparse.
Therefore, in our case, the mesh is created to be more dense in the left side of
the half court, where most shots are located, as illustrated in
Figure~\ref{mesh}. A similar mesh can be found in
\cite{cervone2016multiresolution}.

Estimation of LGCP using INLA is facilitated by the \textsf{R}-package
\textbf{inlabru} \citep{bachl2019inlabru}, which provides an easy access to
Bayesian inference for spatial point processes. A benefit of using
\textbf{inlabru} is that it provides methods for fitting spatial density
surfaces, as well as for prediction, while not requiring knowledge of SPDE
theory. With a mesh created as shown in Figure~\ref{mesh}, the SPDE can be
constructed on the mesh using the function \texttt{inla.spde2.pcmatern()}. The
``pc'' in ``pcmatern'' is short for ``penalized complexity", and it is used to
refer to prior distributions over the hyperparameters that are both
interpretable and have interesting theoretical properties \cite[see,][for a
discussion]{simpson2017penalising}.

Based on INLA, we obtain estimated intensity surfaces
$\hat{\lambda}^{(1)}(\cdot),\hat{\lambda}^{(2)}(\cdot),
\ldots,\hat{\lambda}^{(n)}(\cdot)$, and then obtain $\mathscr{S}$ by
\eqref{eq:k_estimation} and \eqref{eq:sBm}. Next, we use MFM-PPGrouping for
group learning based on $\mathscr{S}$. The sampler presented in
Algorithm~\ref{algorithm1} is used to sample from the posterior distributions
for unknown parameters, including $k$, $z = (z_1, \ldots, z_n) \in \{1, \ldots,
k\}^n$ and $\lambda = (\lambda_1, \ldots, \lambda_n)$ in \eqref{eq:Func_MFM}. As
it marginalizes over the distribution of $k$, the sampler does not depend on
reversible jump, and is more efficient than allocation samplers. 
\begin{algorithm}[tbp]
\caption{Collapsed Sampler for MFM-PPGrouping}
\label{algorithm1}
\begin{algorithmic}[1]
\Procedure{c-MFM-PPGrouping} {}
\\ Initialize: let $\bm{z} = (z_1, \ldots, z_n)$, $\bm{U} = (U_{rs})$,
$\bm{T} = (T_{rs})$.
\For{each iter $=1$ to $\mbox{M}$ }
\\ Update $\bm{T} = (T_{rs})$ conditional on $z$  as
\begin{align*} 
\begin{split}
p(T_{rs} \mid \mathscr{S},z) & \sim \mbox{Gamma}\left(\alpha + n_{rs}/2, \beta
+
(n_{rs}-1)\text{var}(A_{[rs]})/2 + \dfrac{k_0 n_{rs} 
(\bar{A}_{[rs]}-\mu_0)^{2}}{2(k_0 + n_{rs})}\right)
\end{split}     
\end{align*} 
\\ Update $\bm{U} = (U_{rs})$ conditional on $z$ as
\begin{align*} 
\begin{split}
p(U_{rs} \mid \mathscr{S},T_{rs},z) & \sim \mbox{N}\left(\dfrac{k_0 \mu_0 +
n_{rs}
\bar{A}_{[rs]}}{k_0 + n_{rs}}, ((k_0+n_{rs})T_{rs})^{-1}\right)
\end{split}     
\end{align*} 
where ${A}_{[rs]}=(\mathscr{S}_{ij}; {z_i=r,z_j=s,i\neq j})$, $\bar{A}_{[rs]}=
(\sum_{z_i=r,z_j=s,i\neq j}\mathscr{S}_{ij})/n_{rs}$ and $n_{rs} = \sum_{i
\neq
j}I(z_i=r,z_j=s),  r=1,\ldots,k; s=1,\ldots,k$. Note that $k$ denotes the
number of groups yielded by the current $z$.
\\ Update $\bm{z} = (z_1, \ldots, z_n)$ conditional on $\bm{U} = (U_{rs})$ and
$\bm{T} = (T_{rs})$. For each $i$ in $(1,\ldots,n)$,
$P(z_i = c \mid z_{-i}, \mathscr{S}, U, T)$ can be obtained in closed form as:
\[\propto \left\{
\begin{array}{ll}
      [\abs{c} + \gamma] [\prod_{j > i} \frac{1}{\sqrt{2\pi T_{cz_j}^{-1}}}
      e^{-\frac{T_{cz_j}(\mathscr{S}_{ij}-U_{cz_j})^2}{2}}] [\prod_{k < i} 
\frac{1}{\sqrt{2\pi T_{z_kc}^{-1}}} e^{-\frac{T_{z_kc}(\mathscr{S}_{ki}-
U_{z_kc})^2}{2}}] & \text{at an existing table $c$} \\
      \frac{V_n(\abs{\mathcal{C}_{-i}}  +1)}{V_n(\abs{\mathcal{C}_{-i}}}
      \gamma  m(\mathscr{S}_i) & \text{if $c$ is a new table} \\
\end{array},
\right. \]
with $\mathcal{C}_{-i}$ being the partition obtained by removing $z_i$ and 
\begin{eqnarray*}
m(\mathscr{S}_i) =  \prod_{t= 1}^{\abs{\mathcal{C}_{-i}}}
\dfrac{\Gamma(\alpha_n)}{\Gamma(\alpha)} 
\dfrac{\beta^\alpha}{\beta_n^{\alpha_n}} 
\left(\frac{k_0}{k_0+n_t}\right)^{\frac{1}{2}}(2\pi)^{\frac{-n_t}{2}},
\end{eqnarray*}
where $\alpha_n=\alpha + n_t/2$, $n_t = \sum_{i \neq j}I(z_j = t)$,
$\beta_n=\beta + (n_{t}-1)\text{var}(A_{[t]})/2 + \dfrac{k_0 n_{t}
(\bar{A}_{[t]}-\mu_0)^{2}}{2(k_0 + n_{t})}$, ${A}_{[t]}=(\mathscr{S}_{ij};
{z_j=t,i\neq j})$ and $\bar{A}_{[t]}= (\sum_{z_j=t,i\neq
j}\mathscr{S}_{ij})/n_{t}$.

\EndFor
\EndProcedure
\end{algorithmic}
\end{algorithm}

After obtaining posterior samples of $\{z_1,z_2,\ldots,z_n\}$, posterior
inference for the group configurations needs to be carried out so that the
values are nominal integers denoting group belongings. This renders the
posterior mean unsuitable for our purpose. We adopt Dahl's method
\citep{dahl2006model}. Define a membership matrix $\mathcal{B}^{(\ell)}$ as:
\begin{align}
\begin{split}
\mathcal{B}^{(\ell)} = (\mathcal{B}^{(\ell)}(i,j))_{i,j\in \{1:n\}} =
(z_i^{(\ell)}
=
z_j^{(\ell)})_{n\times n},
\end{split}
\end{align}
where $\ell=1,\ldots, B$ indexes the number of post-burnin MCMC iterations,
$z_i^{(\ell)}$ and $z_j^{(\ell)}$ denote the memberships for players $i$ and
$j$, respectively. An entry~$\mathcal{B}^{(\ell)}(i,j)$ equals~1 if
$z_i^{(\ell)}=z_j^{(\ell)}$, and 0 otherwise. An element-wise mean of the
membership matrices can be obtained as
\begin{equation*}
\bar{\mathcal{B}} =
\frac{1}{B}
\sum_{t=1}^B \mathcal{B}^{(t)},
\end{equation*}
where the summation is also element-wise.
The posterior iteration with the smallest squared distance to
$\bar{\mathcal{B}}$
is obtained by
\begin{align}
C_{LS} = \text{argmin}_{c \in (1:B)} \sum_{i=1}^n \sum_{j=1}^n
(\mathcal{B}^{(c)}(i,j) -
\bar{\mathcal{B}}(i,j))^2.
\end{align}  
The estimated parameters, together with the group assignments $\bm{z}$, are
obtained from $C_{LS}$th post burn-in iteration. With the Dahl's method, our
Bayesian grouping method is summarized in Algorithm \ref{algorithm2}.

\begin{algorithm}[tbp]
\caption{Bayesian Group Learning Procedure for Basketball Players}
\label{algorithm2}
\begin{algorithmic}[1]
\item Fit LGCPs for $n$ different players
$\bm{y}^{(1)},\bm{y}^{(2)},\ldots,\bm{y}^{(n)}$ via \textbf{inlabru} and get $n$
underlying intensity surface
$\hat{\lambda}^{(1)}(\cdot),\hat{\lambda}^{(2)}(\cdot),\ldots,\hat{\lambda}^{(n)}(\cdot)$,
\item Use \eqref{eq:k_estimation} and \eqref{eq:fisherZ} to construct matrix
$\bm{S}$ and matrix $\mathscr{S}$ and based on
$\hat{\lambda}^{(1)}(\cdot),
\hat{\lambda}^{(2)}(\cdot),\ldots,\hat{\lambda}^{(n)}(\cdot)$,
\item Get $B$ posterior samples of
$\bm{z}^{(1)},\bm{z}^{(2)},\ldots,\bm{z}^{(B)}$ from $\mathscr{S}$ via Algorithm
\ref{algorithm1},
\item Summary posterior samples by Dahl's method.
\end{algorithmic}
\end{algorithm}

\section{Simulation}\label{sec:simu}
\subsection{Simulation Setup}

\begin{figure}[tbp]
	\centering
	\includegraphics[width = 0.8\textwidth]{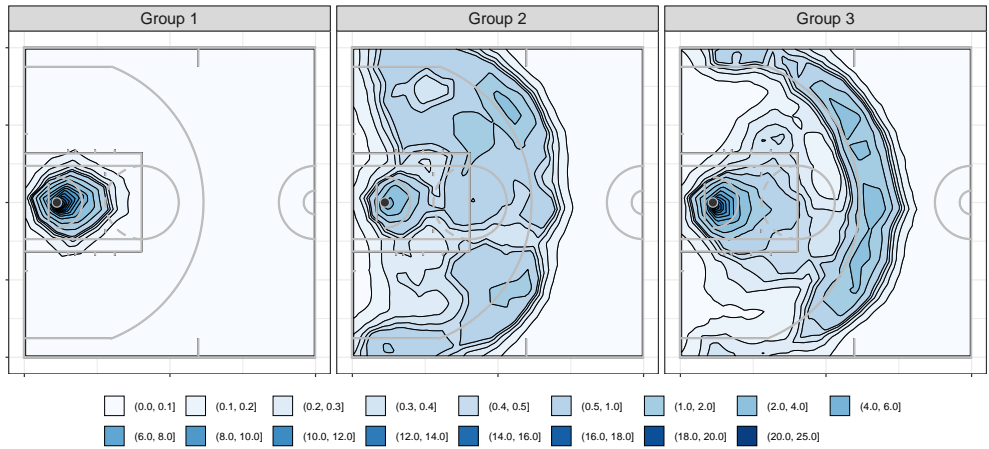}
	\caption{Visualization of three bases for the three groups in simulation
	design.}
	\label{Fig:simulation}
\end{figure}

A total of three groups are designed, each of which has its own base intensity
as shown in Figure~\ref{Fig:simulation}. The first group corresponds to players
most of whose shots are in the painted area; the second group corresponds to
players whose shot locations are widely distributed in every location from
painted area to three-point line. A player in the third group has more shot at
the three-point line and inside the painted area. To create some variation
between players within the same group so that their shot charts are not
generated from exactly the same intensity surface, a noise term has been added
to the base surfaces, so that for player $i$ in group $k$,
\begin{equation}
	\lambda_{k,i}(\cdot) = \lambda_k(\cdot) +
	\vert\bm{\epsilon}_{k,i}\vert,~~k=1,2,3,~~i=1,\ldots, 25,
\end{equation}
where $\lambda_k(\cdot)$ denote the three base intensity surfaces in
Figure~\ref{Fig:simulation}, $\bm{\epsilon}_{k,i}$ is generated from a
multivariate normal distribution with mean~$\bm{0}$ and
variance~$0.5\textbf{I}$, and an absolute value step is taken to ensure that the
summation produces a positive and valid intensity surface.

With 75 valid intensity surfaces having both between group variation and within
group variation, shot locations for the 75 players are generated following the
Poisson process as in Equation~\eqref{eq:spp}. Algorithm~\ref{algorithm2} is
implemented on these 75 players, and we examine both the estimation for number
of groups as well as congruence of group belongings with the true setting in
terms of modulo labeling by Rand index \citep[RI;][]{rand1971objective}, the
computation of which is facilitated by the R-package \textbf{fossil}
\citep{vavrek2011fossil}. The RI ranges from~0 to~1 with a higher value
indicating better agreement between a grouping scheme and the true setting. In
particular, a value of~1 indicates perfect agreement.



\subsection{Simulation Results}
\begin{figure}[tbp]
	\centering
	\includegraphics[width = 0.7\textwidth]{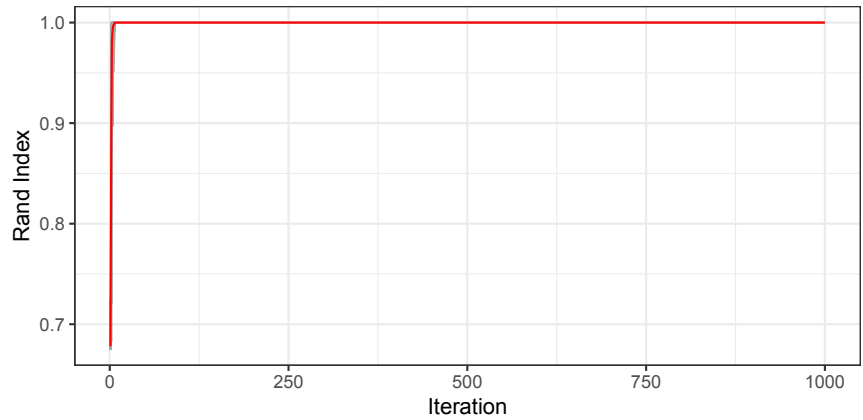}
\caption{Rand Index trace plot for single replicate simulated data. Dark grey
line indicates for each random seed. Red line is the average rand index for 50
random seeds.}
	\label{Fig:simtrace}
\end{figure}

We run our algorithm with 1,000 MCMC iterations, with the first 500 iterations
as burn-in for each replicate data. We examine it is sufficient for the chain to
converge and stabilize. The numbers are chosen to be sufficiently large for the
chain to converge and stabilize. To verify this, with a single replicate of
data, 50 separate MCMC chains are run with different random seeds and hence
initial values, and 50 final grouping schemes are obtained. The RI is calculated
for these 50 chains at each of their iteration, giving 50 traces, which are
visualized in Figure~\ref{Fig:simtrace}. It can be observed that covergence is
attained after a small number of iterations, and the band of the 50 traces is
rather tight after convergence.

\begin{figure}[tbp]
	\centering
	\includegraphics[width = 0.7\textwidth]{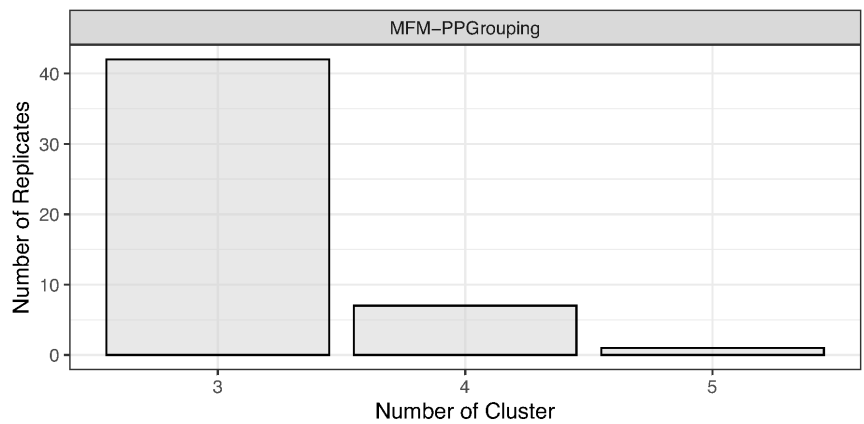}
	\caption{Histogram for the number of clusters produced by MFM-PPGrouping
	in 50 simulation replicates.}
	\label{Fig:simcluster}
\end{figure}

Proceeding to 50 separate replicates of data, our proposed algorithm was run,
and~50 RI values are obtained by comparing with the true setting. They average
to 0.9988, which indicates rather accurate grouping ability of the proposed
approach. In addition, performance comparisons of our proposed method with three
competing methods are made. We compare our method to K-means algorithm,
Density-based spatial grouping of applications with noise (DBSCAN) and mean
shift grouping. Grouping recovery performances of all four methods are measured
using the RI. The~50 final RI's obtained for the three competitors average to
0.9005, 0.7642, and 0.7380, respectively, indicating the superior performance of
our proposed approach. We also show the number of cluster covered by our
algorithm, see Figure~\ref{Fig:simcluster}. We see that there are fourty two
replicates have three clusters.

\section{Analysis of NBA Players}\label{sec:real_data}

\begin{figure}[thp]
	\centering
	\includegraphics[width = \textwidth]{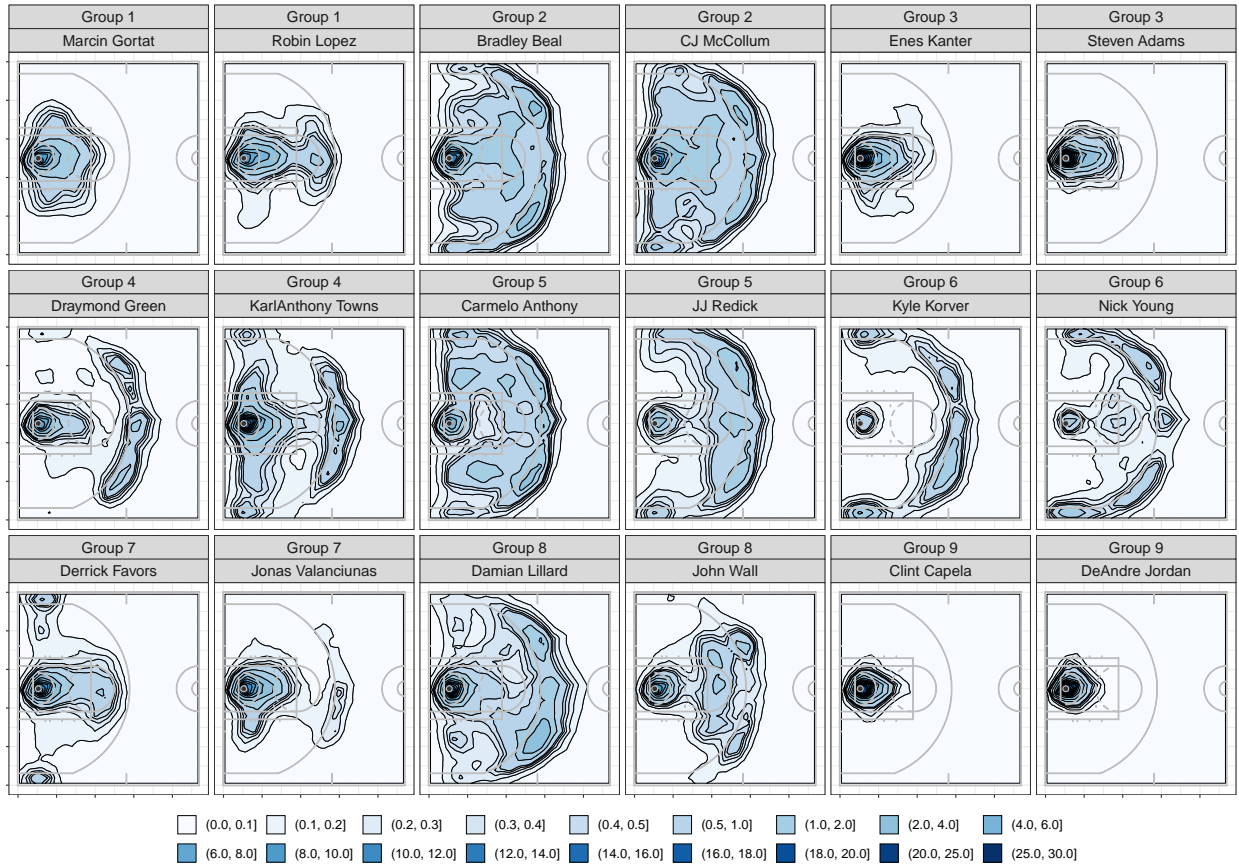}
	\caption{Fitted intensities with contour lines
	for two selected players in each of the nine 
	identified groups.}
	\label{Fig:SelectedPlayersCluster}
\end{figure}

In this section, we apply the proposed method to the analysis of players' shot
data in the 2017-2018 NBA regular season. Only the locations of shots are
considered regardless of the players' positions on the court (e.g., point guard,
power forward, etc.). Players' positions (e.g., point guard, power forward, etc)
are not considered, but only the shots' positions. As a starting point, a
predictive intensity matrix is obtained for each player using \textbf{inlabru}.
Algorithms~\ref{algorithm1} and~\ref{algorithm2} are subsequently used to
identify the groups. We run 1,000 MCMC iterations and the first 500 iterations
as burn-in period. The result from the MFM model suggests that the 191 players
are to be classified into nine groups. The sizes of the nine groups are 10, 59,
8, 19, 24, 8, 10, 51, and 2 respectively. Visualizations of intensity matrices
with contour for two selected players from each group are presented in
Figure~\ref{Fig:SelectedPlayersCluster}.



Several interesting observations can be made from the visualization results.
First, we discuss groups~1, 3 and 9. The contours for group~1 and group~3 are
wider than the contours for group~9, where most shots are located near the hoop.
Clint Capela and DeAndre Jordan (in group~9), for example, are both good at
making alley-oops and slam dunks. Only very few shots are made by these players
outside the painted area. Despite the similarity between groups~1 and~3, it can
be seen that for group~3, most of the shots are made within the painted area,
while there are quite a number of shots outside the painted area for group~1,
indicating wider shooting ranges for the corresponding players.

Groups~4 and~7 share some characteristic in common as most of the shot locations
are around the hoop. Players in group~4, however, are also able to shoot
frequently beyond the three-point line at a wider range of angles, while players
in group~7 also shoot beyond the three-point line at very limited angles. 


Groups~2, 5 and~8 also bear some resemblance with each other. A first look at
the fitted intensity contours indicates that players in these groups are able to
make all types of shots, including three-pointers, perimeter shot, and also
shots over the painted area. Group~2 is differentiated from the other two as
most of the shots are made around the hoop, and the intensity for perimeter
shots and three-pointers are rather similar. For group~5, however, the intensity
for shots around the hoop is much lower than that for group~2. For group~8,
players make most shots around the hoop and also make some perimeter shots as
well as three-pointers. Compared with the other two groups, they have a more
narrow range of angles to make perimeter shots and three-pointers. Most shots
are located between 45 degrees up and down angle from the horizontal line across
the hoop.

Group~6 has little similarity with any other groups. As can be seen, most shots
are located either near the hoop, or beyond the three-point line. There are very
few perimeter shots. Kyle Korver and Nick Young, both of whom are well-known
catch-and-release shooters, fall in this group.

\begin{figure}[thp]
	\centering
	\includegraphics[width=0.8\textwidth]{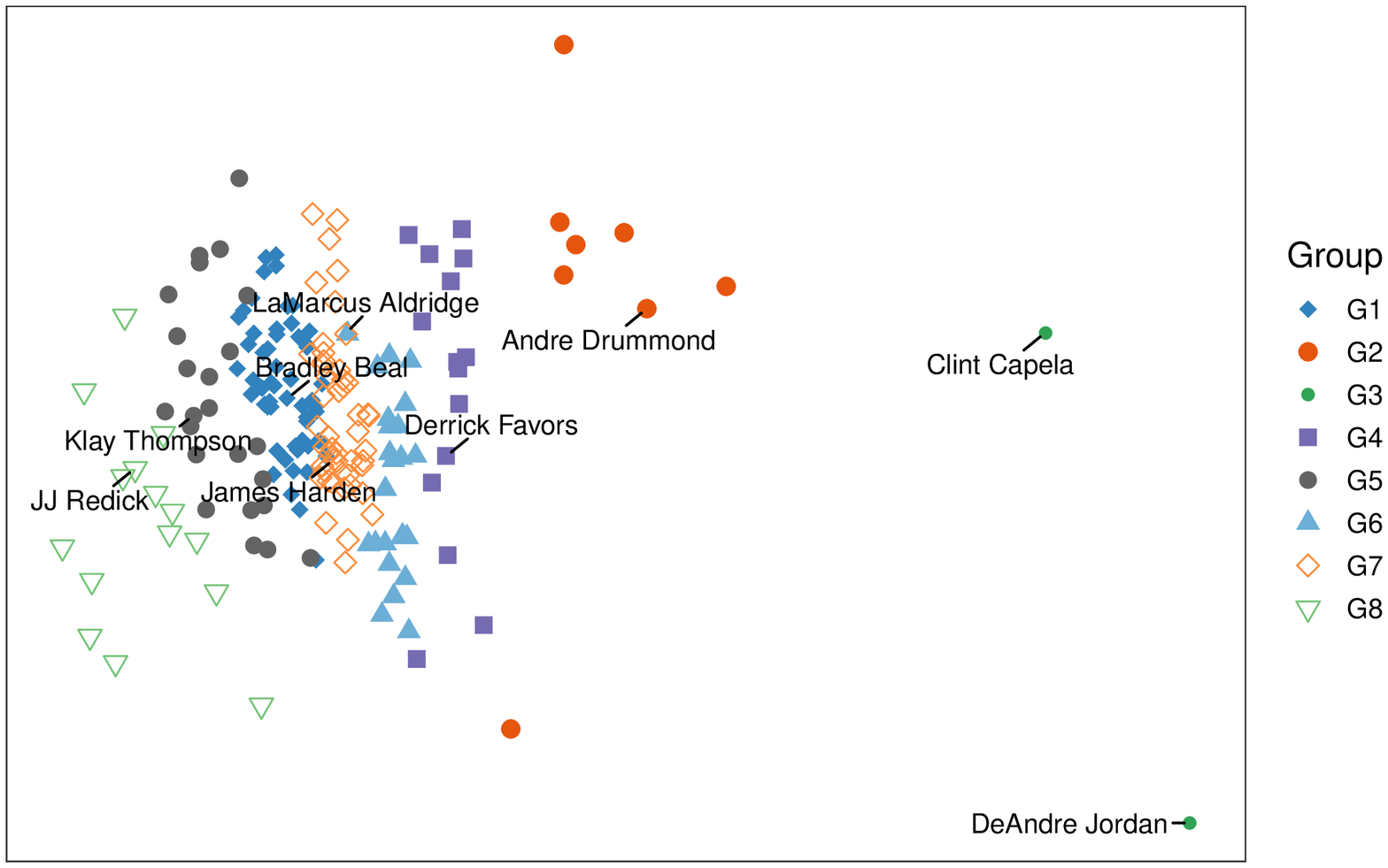}
	\caption{Visualization of the night identified groups of players, with
selected
	players' names annotated.}
	\label{fig:cluster}
\end{figure}

As further verification, we use multidimensional scaling to lower the dimension
of the fitted intensity matrices for players to 2 so that similarities in their
shooting habits can be visualized. See Figure~\ref{fig:cluster}. Separation of
the nine groups is quite clear. Group~9, for example, with its unique strong
preference for alley-oops and slam dunks, stands far from others.

Finally, to make sure the group configuration presented here is not a random
occurrence but reflects the true pattern demonstrated by the data, we run 50
separate MCMC chains with different random seeds and initial values, and
obtained 50 final grouping schemes. The RI between each scheme and the present
grouping scheme is calculated, and they average to 0.948, indicating high
concordance of conclusion regardless of random seeds.

\section{Conclusion}\label{sec:discussion} In this paper, we proposed using MFM
to capture heterogeneity of different NBA players based on LGCP. Our group
learning method provides a quantitative summary of different players shot habits
other than traditional position categorization. Our simulation results indicated
that our proposed methods achieve good grouping accuracy.

The real data application give us the information about player's shooting habit
location. Players can understand their own shooting habits, and they can also
strengthen their weaker shooting locations. On the other hand, the professional
coach can formulate a defensive strategy to reduce the opponent's score with
these information. Our grouping results will provide a good guidance for team
managers trading the players with similar shot pattern.

A few topics beyond the scope of this paper are worth further investigation. In
this paper, a two-stage group learning method is proposed. An unified approach
is an interesting alternative in future work. In addition, incorporating
auxiliary information such as player position or historical information could
also be taken into account for grouping in our future work. Jointly modeling
spatial field goal percentage and shot selection will provide more detail
instructions for professional coaches.

\end{document}